\documentclass[letterpaper]{article} %
\usepackage[]{aaai24}  %
\usepackage{times}  %
\usepackage{helvet}  %
\usepackage{courier}  %
\usepackage[hyphens]{url}  %
\usepackage{subfigure}
\usepackage{graphicx} %
\usepackage{natbib}  %
\usepackage{caption} %
\usepackage{algorithm}
\usepackage{algorithmic}

\usepackage{newfloat}
\usepackage{listings}
\DeclareCaptionStyle{ruled}{labelfont=normalfont,labelsep=colon,strut=off} %
\floatstyle{ruled}
\newfloat{listing}{tb}{lst}{}
\floatname{listing}{Listing}
\newcommand{\descr}[1]{\smallskip\noindent\textbf{#1}}

\newcommand{\weburlfootnote}[1]{%
	\stepcounter{footnote}%
	\footnotemark[\thefootnote]%
	\footnotetext[\thefootnote]{\url{#1}}%
}

\usepackage{afterpage}
\usepackage[bottom]{footmisc}

\title{Gun Culture in Fringe Social Media}
\author{
	Fatemeh Tahmasbi\textsuperscript{{\rm 1}},
	Aakarsha Chug\textsuperscript{{\rm 1}},
	Barry Bradlyn\textsuperscript{{\rm 2}},
	Jeremy Blackburn\textsuperscript{{\rm 1}}
}
\affiliations{
	\textsuperscript{\rm 1}Binghamton University,
	\textsuperscript{\rm 2}University of Illinois Urbana-Champaign\\
	ftahmas@binghamton.edu, alnu2@binghamton.edu, bbradlyn@illinois.edu, jblackbu@binghamton.edu
}

\begin{document}
	
	\maketitle
	
	\begin{abstract}
		The increasing frequency of mass shootings in the United States has, unfortunately, become a norm. 
		While the issue of gun control in the US involves complex legal concerns, there are also societal issues at play.
		One such social issue is so-called ``gun culture,'' i.e., a general set of beliefs and actions related to gun ownership.
		However relatively little is known about gun culture, and even less is known when it comes to fringe online communities. This is especially worrying considering the aforementioned rise in mass shootings and numerous instances of shooters being radicalized online.
		
		To address this gap, we explore gun culture on /k/, 4chan's weapons board.
		More specifically, using a variety of quantitative techniques, we examine over 4M posts on /k/ and position their discussion within the larger body of theoretical understanding of gun culture.
		Among other things, our findings suggest that gun culture on /k/ covers a relatively diverse set of topics (with a particular focus on legal discussion), some of which are signals of fetishism.
	\end{abstract}

	\section{Introduction}

	The United States is relatively unique in the world when it comes to gun ownership.
	In fact, \emph{gun culture}, i.e., the set of beliefs and activities oriented around the ownership and use of guns~\cite{witkowski2014visual}, has been at the heart of America since its founding~\cite{yamane2017sociology}.
	Gun culture in the US is entrenched in politics, religion, and all facets of society in between~\cite{gun_politics_religion} to the point that it can turn into \emph{fetishism}~\cite{texasfirearms22}.
	More recently, there has been a large amount of debate about guns due to the increasing number of mass shooting events in the US, and gun culture makes up a major component of discussion~\cite{deangelis2018mass}.

	Consider the 2022 Buffalo shooting, where the shooter purchased a Bushmaster XM-15 rifle in Endicott, NY and murdered 10 people in a grocery store in a black neighborhood.
	The shooter---a self-described right-wing extremist---was in large part radicalized via social media, in particular 4chan~\cite{nbc_buffalo_shooter_manifesto}.
	While 4chan's influence in terms of both online and real-world extremism is relatively well studied~\cite{zannettouQuantitativeApproachUnderstanding2020,4chan,right_wing_extremist}, its position in terms of \emph{gun culture} remains murky.
	This is particularly problematic considering that the Buffalo shooter made use of 4chan's weapons board, /k/, to discuss, educate himself on, and even purchase firearms.

	To address this gap, in this paper we explore gun culture on 4chan's /k/, with a specific focus on the intersection of gun culture and fetishism.
	We answer the following research questions:
	\begin{itemize}
		\item RQ1. How does gun culture manifest itself in users' discussions, and to what extent?
		\item RQ2. Are there any signs of gun attachment among users' discussions? 
		\item RQ3. Are there any indications of gun fetishism in users' discussions?
		\item RQ4. Is it possible to identify whether a post reflects gun fetishism?
	\end{itemize}

	To answer these research questions we collect over 4M posts from 4chan's /k/ board.
	We develop a topic analysis pipeline to explore the primary subjects discussed by users on /k/.
	Then we use language modeling techniques to explore different types of attachment to guns in these discussions.
	Finally, we develop a semantic similarity pipeline to identify and analyze posts with signs of gun fetishism.

	Our contributions are as follows:
	\begin{itemize}
		\item Through topic analysis, we observe discussions around two distinct branches of gun culter, termed Gun Culture 1.0 and 2.0. Gun Culture 1.0 typically associates firearms with traditional values, such as masculinity and sentimental values, while Gun Culture 2.0 emphasizes self-defense, empowerment, and ideals like freedom and patriotism~\cite{yamane2017sociology}. We find discussions on /k/ place a greater emphasis on topics associated with Gun Culture 2.0.
		\item We conducted in-depth semantic similarity analysis to explore various types of gun attachments and their contexts in Gun Cultures 1.0 and 2.0, revealing insights into users' language usage regarding these attachments.
		\item We investigate signs of gun fetishism through semantic similarity search, presenting examples of user posts that indicate the potential presence of gun fetishism.
	\end{itemize}

	\section{Background and Related Work}

	\subsection{Gun Culture in America}
	Gun violence in the US is a severe concern, with death rates over 20 times higher than in other countries~\cite{GRINSHTEYN2016266}. In 2019 there were 3.96 shooting deaths per 100,000 people in the US, 8 times higher than in Canada and 100 times higher than in the UK~\cite{us_gun_violence_rate}.
	In 2018, 42\% of American households owned guns, with two-thirds owning multiple firearms~\cite{enten2018there}.
	In 2016, gun-related deaths reached 38,000~\cite{lopez2018america}.
	Several studies explore gun violence~\cite{wamser2021understanding,cukier2018gunViolence,bright2022problem}, and a link between gun violence and US gun culture is recognized~\cite{lemieux2014effect,mencken2019gun}.
	Witkowski~\cite{witkowski2014visual} defines gun culture as beliefs and practices around firearm ownership, leading to varied interpretations among gun owners~\cite{wamser2021understanding}.
	Gun culture is quite diverse~\cite{wamser2021understanding} and is a contributor to gun violence, shaping attitudes and actions towards firearms~\cite{america_gun_culture}.

	 Gun culture in the US is predominantly male-centered, emphasizing traits like dominance, toughness, and aggression~\cite{king2007arming}. This orientation has even led to the sexualization of firearms, as seen in slang equating guns to male anatomy (meat pistol, love gun, etc.)~\cite{king2007arming}. Furthermore, guns are recognized as social facilitators, with some arguing that they act as ``social glue,'' binding gun owners into a collective identity~\cite{sarat2019both}. In certain states, firearms are integral to core identity, exemplified by phrases like ``We are Texas because of guns''~\cite{texasfirearms22}.
	 
	Gun culture is categorized into Gun Culture 1.0 and 2.0~\cite{yamane2017sociology,texasfirearms22}. In Gun Culture 1.0, firearms serve roles in colonization, sports shooting, and hunting, also holding value as collectibles. Firearms often signify masculinity linked with specific religious practices. This is exemplified when boys receive guns during puberty as symbolic gestures. In Gun Culture 1.0, the National Rifle Association (NRA) has actively endorsed firearms through marksmanship competitions~\cite{texasfirearms22,littlefield2011socialization,kalesan2016gun}.

	Gun Culture 2.0 shifts focus to guns as imperative for self-defense~\cite{yamane2017sociology}.
	This transformation is evidenced by, e.g., the NRA advocating for gun ownership by employing succinct slogans like ``guns don't kill people, people kill people''~\cite{henigan2016guns}.
	This sentiment becomes closely associated with the notion of the Second Amendment's perceived ``God-Given'' right to confront ``evil'' forces~\cite{texasfirearms22,yamane2017sociology}.
	Gun owners' attachment to their firearms stems from diverse motivations, rooted in the emotions and significance associated with guns~\cite{mencken2019gun, yamane2017sociology,witkowski2014visual}.

	In Gun Culture 1.0, guns provided leisure and are associated with concepts like masculinity, evoking emotions~\cite{yamane2017sociology}. 
	Conversely, in Gun Culture 2.0, guns are essential for self-defense, imparting both physical and moral empowerment.
	They symbolize ideals like freedom, patriotism, and heroism, often intertwined with religious beliefs~\cite{yamane2017sociology}.
	This infusion of religious belief can be seen as a type of \emph{fetishism}~\cite{texasfirearms22}.
	Fetishism represents widely held yet false beliefs within groups, manifesting in various ways, either directly or indirectly~\cite{kaplan2006cultures}.
	Some of these manifestations include:
	\begin{itemize}
		\item Any activity that is pursued with a heightened sense of urgency.
		\item Belief in magical powers of objects to protect their owner.
		\item The object being inhabited by a god or spirit, determining its worshiper's fate.
		\item Redirection of sexual interest to an object.
		\item Devotion to religious or cultural practices marked by magical fetishes.
	\end{itemize}
	The transition from Gun Culture 1.0 to 2.0 is a contributor to the overall fetishization of firearms~\cite{witkowski2014visual}.

	\subsection{Gun Culture on Social Media}

	To the best of our knowledge, no comprehensive study has examined gun culture and fetishism online at scale.
	However, existing research explores users' views on various gun and mass shooting topics.
	
	Some work investigates online attention towards mass shootings.
	For example, it has been shown that mass shooting events amplify gun control discussions, particularly in their aftermath~\cite{garimella2017effect}.
	Another analysis of  attention across political subjects including abortion, gun control, and Black Lives Matter on YouTube and Twitter, established interconnections through video hyperlinks analysis~\cite{politicalTopicsAttention22} .
	In a general sense, the gun control debate has been studied in the context of support and opposition on Twitter~\cite{benton2016after}.
	The impact of mass shootings on information-seeking behaviors on gun control, as well as alternative narratives about them have also been explored~\cite{koutra2015events,starbird2017examining}.
	Demographic features have also shown to be significant in online gun debate discourse and correlate with likelihood of participation in real-world marches~\cite{mejova2022modeling}.

	There have also been small-scale analyses of gun related image sharing on social media.
	560 images posted to Twitter by black youth gangs in Chicago have been used to understand the differences in how domain experts and social workers differ in their understanding and interpretation of gun culture in gangs~\cite{chicago_youth}.
	2,680 images posted by major gun related publications to Instagram were also explored in the development of a so-called Gun Culture 3.0~\cite{gun_culture3}. However, there have been no analysis studies examining the large-scale landscape of gun culture on social media.

	The remainder of this paper explores 4chan's /k/ within the theoretical framework of gun culture and fetishism described above. 
	However, we do note that the line between obsession and fetishism is both thin and blurry, and distinguishing them is beyond the scope of this paper.
	
	\section{Dataset}
	
	\begin{table}[t]
		\centering
		\begin{tabular}{l c l c }
			\textbf{Threads} & \textbf{Posts} & \textbf{Start Date} & \textbf{End Date}
			\\
			\hline
			77,366 &   4,431,706 & 01/01/2022 & 12/31/2022
			\\
			\hline
			
		\end{tabular}
		\caption{Dataset details.}
		\label{tbl:dataset}
	\end{table}
	
	This paper studies 4chan, a fringe social media platform and image board.
	Threads are created by an \emph{original poster} (OP) with an image, and users reply in flat threads.
	4chan supports anonymity, with users posting as ``Anonymous,'' with no account creation required.
	It is also ephemeral, as threads are deleted after several days at most~\cite{keks}.
	4chan is broken up into \emph{boards}, each focusing on specific topics.
	4chan has a history of hosting toxic and extremist communities~\cite{manoel_manosphere}, and some mass shooters claimed inspiration from the site~\cite{nbc_buffalo_shooter_manifesto}.
	
	We collected the data from all posts on 4chan's weapons board, /k/, in 2022. 
	Discussions on /k/ primarily revolve around firearms, but also cover military and various other weapon related topics.
	Frequent users of this board are colloquially known as ``/k/ommandos''\weburlfootnote{https://knowyourmeme.com/memes/sites/k--4}.
	Our dataset contains over 4.4M posts, detailed in Table~\ref{tbl:dataset}.
	
	\descr{Content Warning.}
	The dataset we use in this paper is from a notoriously toxic community.
	We do not censor any of the content drawn directly from posts in our dataset, but we warn the reader that this paper contains content that is likely to be considered offensive and disturbing.
	
	\section{Content Analysis}
	
	\begin{figure}[t]
		\centering
		\subfigure{\includegraphics[width=0.9\columnwidth]{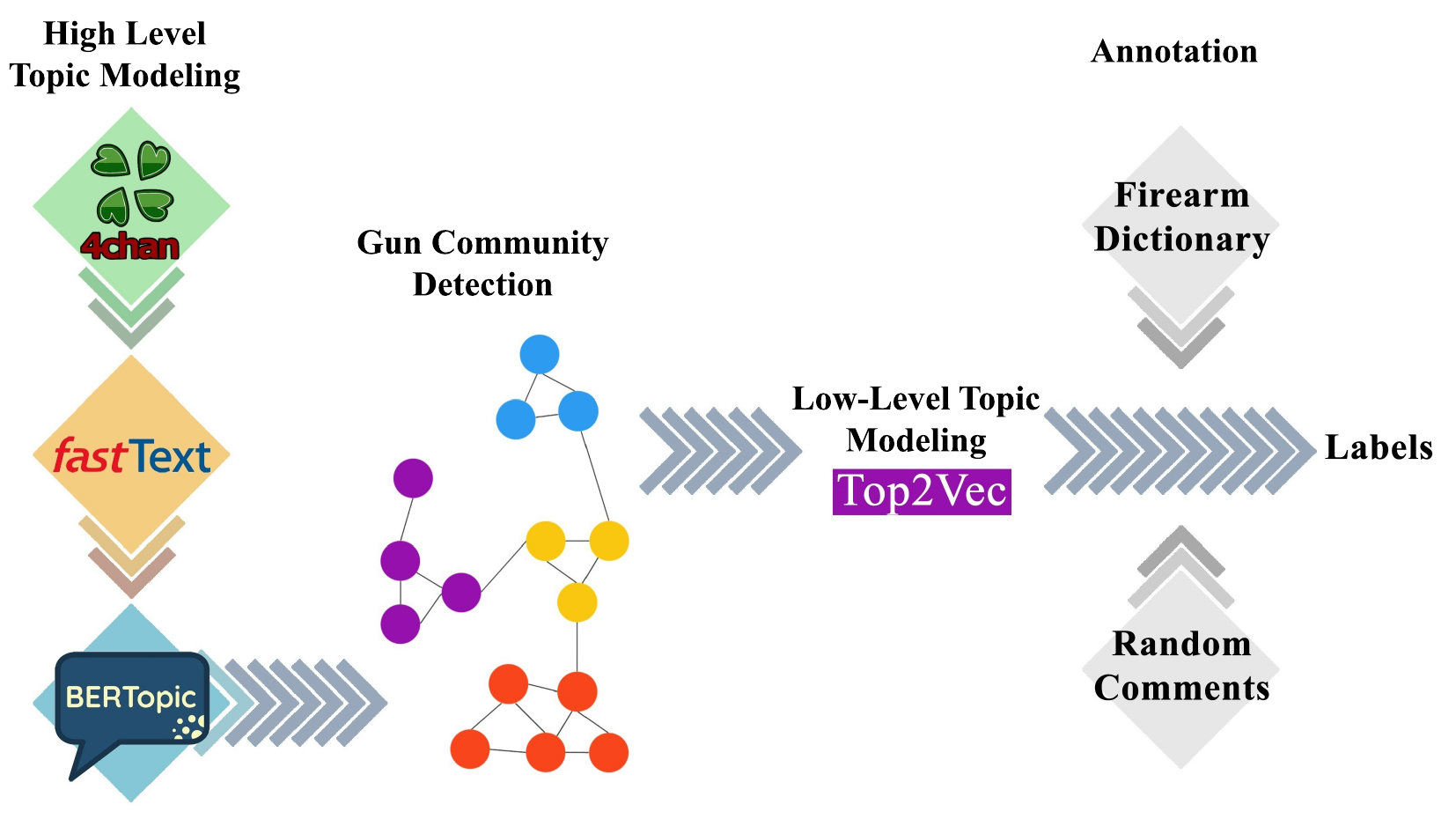}}
		\caption{Gun culture topic analysis pipeline.}
		\label{fig:diagram}
	\end{figure}
	
	\begin{figure*}[t]
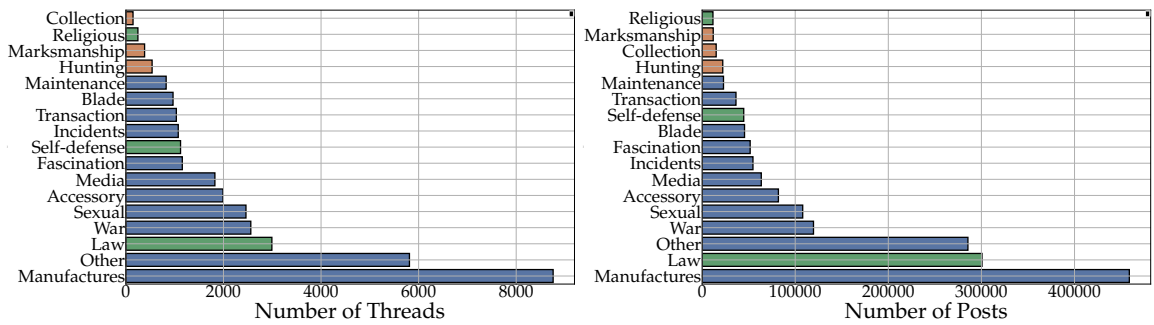

		\centering
		\subfigure{\includegraphics[width=0.9\columnwidth]{figures/k_threads_labels_hist_4.pdf}}
		\subfigure{\includegraphics[width=0.9\columnwidth]{figures/k_comments_labels_hist_4.pdf}}
		\caption{Number of threads and posts related to guns within each discussion labels. Labels directly aligned with Gun Culture 1.0 are highlighted in orange, those directly associated with Gun Culture 2.0 are green, and other labels are blue.}
		\label{fig:k_codebook_frequency}
	\end{figure*}
	
	\begin{table}[t]
		\small
		\begin{tabular}{|l| p{0.5\linewidth}| p{0.25\linewidth} |}
			\hline
			{Cluster \#}    &BERTopic topics & Topic  \\ 
			\hline
			1  &
			we\_russian\_russia\_vdv, kherson\_russians\_russian\_ukraine, nuclear\_nukes\_nuke\_russia, russia\_ukraine\_crimea\_putin, etc.
			&      Russia-Ukraine war  \\
			\hline
			2 &
			gun\_9mm\_glock\_pistol,
			rifle\_sig\_rifles\_m14,
			glock\_trigger\_slide\_beretta,
			ak\_aks\_wbp\_wasr, dot\_optic\_eotech\_holosun,
			oil\_grease\_lube\_rust, shotgun\_shotguns, etc.
			& Gun-related \\
			\hline
			3 &
			artillery\_rocket\_rockets\_guided, radar\_missile\_radars\_missiles, subs\_ships\_navy\_submarine, tank\_tanks\_turret\_armor, etc.	
			& Warfare-related\\
			\hline

		\end{tabular}
		\caption{BERTopic examples across different clusters.}
		\label{table:communities}
	\end{table}

	\subsection{Topic Analysis}
	
	We begin by answering our first research question: \emph{How does gun culture manifest itself in users' discussion, and to what extent?}
	To do this, we build a topic analysis pipeline, a high-level overview of which can be seen in Figure~\ref{fig:diagram}.
	Our pipeline consists of four stages that allow us to measure and understand users' discussions on /k/: 1)~High Level Topic Modeling, 2)~Gun Community Detection, 
	3)~Low Level Topic Modeling, and 4)~Manual Annotation, each of which we describe in more detail below.

	\descr{Stage 1: High Level Topic Modeling}
	In the first stage, we apply BERTopic to the threads in our dataset.
	BERTopic is an NLP technique that uses the BERT language model to cluster documents and generates topics for each cluster based on the TF-IDF procedure~\cite{bertopic}. For details on BERTopic modeling please refer to the Appendix.
	BERTopic ultimately generates 377 topics.

	\descr{Stage 2: Gun Community Detection.}
	In the second stage, we use a community detection algorithm to partition these topics into separate clusters by constructing a weighted graph where topics are represented as nodes, and edge weights are determined by cosine similarity using BERTopic embeddings.
	The Louvain Community detection algorithm~\cite{louvian} is subsequently applied to this graph, identifying three distinct clusters. Examples of randomly selected topics within each cluster can be found in Table~\ref{table:communities}.
	Following manual inspection, we choose the cluster with topics primarily focused on guns and shootings (cluster 2 in Table~\ref{table:communities}) for further analysis, while the other clusters revolve around warfare and the Russia-Ukraine conflict.
	All threads within the selected cluster are then retrieved.

	\descr{Stage 3: Low Level Topic Modeling.}
	In the third stage, we use Top2Vec to identify topics within the chosen gun-related threads from the second stage.
	We choose Top2Vec here for its capacity to generate more coherent and interpretable topic representations, as it takes into account all words near the cluster's centroid, assuming that document and word embeddings share the same vector space~\cite{bertopic,top2vec}. 
	This results in more detailed and easily interpretable topic representations for our use case.

	\descr{Stage 4: Manual Annotation.}
	In the fourth and final stage of our pipeline we manually annotate the 198 quantitatively derived topics from the previous stage.
	We build a gun topic dictionary to serve as a reference guide for understanding gun-related topics, phrases, and abbreviations, enabling more accurate labeling.
	Our gun topic dictionary is based on both an online firearm dictionary~\cite{gun_dictionary} as well as manual inspection and discussion between authors of the paper to handle out-of-dictionary phrases and words (e.g., slang).
	After developing the dictionary, three authors of the paper annotated the topics over three iterations with the goal of extracting the central theme of discussion.
	In the end, our annotators reach a Fleiss' Kappa score of 0.75, suggesting \emph{substantial inter-annotator agreement}~\cite{fleiss1981measurement}.

	\descr{Results.}
	Below is the list of labels and their corresponding inspirations based on gun  discussions in related works:
    
	\begin{itemize}
		\item \descr{Law.} Discussions related to various laws and political parties' approach towards guns and gun ownership~\cite{texasfirearms22, witkowski2014visual, yamane2017sociology}.
		
		\item \descr{Sexual.} Posts related to guns in a sexual context or sexual topics being discussed in gun-related threads~\cite{texasfirearms22, witkowski2014visual, king2007arming}.
		\item \descr{Hunting.} Discussions that are mostly about hunting~\cite{yamane2017sociology}.
		\item \descr{Collection.} Posts that view guns as valuable objects, both sentimentally and financially, as well as the collection, gifting, and inheritance of guns~\cite{yamane2017sociology}.
		\item \descr{Fascination.} Discussions related to people's fascination with different gun features, like shooting power and the smell of bullets~\cite{witkowski2014visual}.
		 It can also be a form of fetishism towards guns~\cite{witkowski2014visual}.
		\item \descr{Manufacturers.} Discussions about different gun manufacturers.
		Gun culture symbolizes guns and weapons as sources of power and superiority, and gun manufacturers play a central role in perpetuating this culture~\cite{luckham1984armament}.
		\item \descr{Maintenance.} Discussions about different gun maintenance tools and skills~\cite{texasfirearms22}.
		\item \descr{Transaction.} Discussions about purchasing guns and the processes involved in buying them~\cite{witkowski2014visual}.
		\item \descr{Marksmanship.} Discussions related to improving shooting skills~\cite{yamane2017sociology}.
		\item \descr{Self-defense.} Discussions about using guns as a tool for self-defense to protect oneself, family, property, and one's country and freedom~\cite{yamane2017sociology}.
		\item \descr{Religious.} Discussions related to gun ownership being a God-given right or the introduction of religion into gun-related discussions, as discussed in~\cite{texasfirearms22}.
		\item \descr{Media.} Discussions about gun-related topics in different types of media, including books, movies, and video games.
		Gun culture can manifest itself in these outlets by fetishizing guns through making heroes and heroism associated with guns, as seen in movies and comics~\cite{captainAmerica}.
		\item \descr{Accessories.} Posts and discussions about different gun and shooting accessories~\cite{hunting_accessory}.
		\item \descr{Incidents.} Posts related to mass shooting incidents~\cite{gme2023impact}.
	\end{itemize}
	
	The remaining labels are derived from our analysis of the topics and discussions.
	While not directly derived from the theoretical literature, they are nevertheless distinct topics discussed on /k/, and use useful to getting a better understanding of gun culture and fetishism:
	\begin{itemize}	
		\item \descr{Blades.} Discussions about other weapons, like blades and swords.
		\item \descr{War-related.} Discussions about guns in a wartime context.
		\item \descr{Other.} This label is applied to topics that did not fit into any of the previously mentioned categories. It is important to note that posts in this label may also reflect one or more of the aforementioned labels alongside other random discussions. Posts in these threads are not sufficiently cohesive to be assigned to other labels.
	\end{itemize}
	
	Figure~\ref{fig:k_codebook_frequency} plots the number of threads and posts related to each label. Labels directly associated with Gun Culture 1.0 are shown in orange and labels directly reflecting Gun Culture 2.0 are shown in green.

	\descr{Takeaways.}
	On /k/, users discuss elements from both Gun Culture 1.0 and 2.0, with ``Manufacturers'' being the most prevalent label.
	In these threads, users primarily delve into various gun types, their manufacturers, and quality. Gun Culture 1.0 labels like ``Collection,'' ``Marksmanship,'' and ``Hunting'' are less frequently discussed in comparison to Gun Culture 2.0 concepts like ``Law,'' ``Self-defense,'' and ``Religion.'' It is notable that discussions around ``Sexual'' topics are highly prevalent, possibly indicative of gun fetishism.
	
	\subsection {Gun Attachment and Semantic Analysis}

	We conduct a detailed analysis of language usage on /k/ to delve into gun culture, specifically focusing on aspects of gun attachment. We examine motivations rooted in emotions and significance associated with firearms, drawing from theoretical literature to explore the diverse perceptions of guns in Gun Culture 1.0 and Gun Culture 2.0.
	We consider expressions of gun fetishism (i.e., attachment or obsession) within the dimensions of gun culture, examining how these perceptions manifest in users' posts.
	Drawing from Kaplan's definition~\cite{kaplan2006cultures}, gun fetishism can be expressed as:
	\begin{itemize}
		\item Viewing guns as magical objects.
		\item Considering guns as religious objects. 
		\item Replacing a higher power with guns.
		\item Perceiving gun possession as a necessity.
		\item Associating guns with sexual and erotic desires.
		\item Placing sentimental values on guns.
		\item Attributing social value (group identity) and self-worth to guns.
	\end{itemize}

	To examine gun attachment manifested in various forms within user discussions and comments, we use Word2Vec, a lightweight language model with keyword search functionality~\cite{word2vec}.
	By training the model on cleaned and stemmed posts from our pipeline's stage 2, we can gain insights into the context surrounding gun culture keywords.
	This approach---which has been used in the past to understand aspects of language of cryptic/niche online communities, including 4chan~\cite{papasavvaItQoincidenceExploratory2021,tahmasbiGoEatBat2021,zannettouQuantitativeApproachUnderstanding2020}---not only reveals whether users discuss these concepts, but also helps us understand them in context of the overall discussion around any particular set of keywords.
	We focus on keywords inspired by existing works and those identified by the model, providing a nuanced exploration of different aspects of gun culture, attachment, and potential signs of fetishism among users.
	While creating a comprehensive keyword list is impractical, the model helps by suggesting similar words.
	See the Appendix for details on data cleaning and model parameters.

	\begin{table}[t]
		\centering
		\begin{tabular}{l c | l c }
			\hline
			\textbf{Key Word} & \textbf{Cosine} & \textbf{Key Word} & \textbf{Cosine} \\
			(gun+collect) & \textbf{Sim.}&  (gun+friend\_famili)&\textbf{Sim.} \\
			\hline
			collector & 0.69 & commun & 0.65  \\
			firearm & 0.59 & peopl & 0.64 \\
			safe\_queen & 0.574 & encourag & 0.62 \\
			hobbi & 0.56 & gun\_owner & 0.62 \\
			antiqu & 0.55 & stranger & 0.59 \\
			milsurp & 0.52 &  gun\_ownership & 0.59 \\
			novelti & 0.51 & social\_circl & 0.58 \\
			sentiment\_valu & 0.50 & relationship & 0.58 \\
			handgun & 0.50 & hobbi & 0.57 \\
			heirloom & 0.49 & motiv & 0.55 \\
			\hline
		\end{tabular}
		\caption{Top 10 similar words to the gun collection and social-related keywords resulting from from Word2Vec.}
		\label{tab:collection_social}
	\end{table}
	
	\begin{table*}[t]
		\centering
		\begin{tabular}{l c | l c | l c}
			\hline
			\textbf{Key Word} & \textbf{Cosine} & \textbf{Key Word} & \textbf{Cosine} & \textbf{Key Word} & \textbf{Cosine} \\
			(marksmanship) &\textbf{Sim.} &  (hunt) &\textbf{Sim.}& (trophi\_hunt)& \textbf{Sim.} \\
			\hline
			profici & 0.76 & hunter & 0.82 & nearli\_extinct & 0.80 \\
			disciplin & 0.71 & hunt\_deer & 0.81 & bear\_wolv & 0.79  \\
			marksman & 0.70 & small\_game & 0.80 & fox\_coyot & 0.77 \\
			becom\_profici & 0.66 & varmint & 0.74 & hummingbird & 0.77  \\
			competit\_shooter & 0.66 &hunt\_bird & 0.72 & safari\_hunt & 0.76  \\
			skill & 0.64 & danger\_game & 0.71 & husbando & 0.75 \\
			basic\_marksmanship & 0.64 & whitetail & 0.70& lead\_gender & 0.75  \\
			practic & 0.64 & pest\_control & 0.70 & happi\_trail & 0.75 \\
			train & 0.64 & hunt\_season & 0.67 & hunting &  0.75 \\
			long\_rang & 0.61 & medium\_game & 0.67& mah\_nigga & 0.75 \\
			\hline
		\end{tabular}
		\caption{Top 10 similar words to the marksmanship and hunting-related keywords resulted from Word2Vec.}
		\label{tab:marksmanship}
	\end{table*}
	
	\begin{table}[t]
		\centering
		\begin{tabular}{l c | l c }
			\hline
			\textbf{Key Word} & \textbf{Cosine} & \textbf{Key Word} & \textbf{Cosine} \\
			(gun+religion) & \textbf{Sim.}& (bear\_arm)&\textbf{Sim.}  \\
			\hline
			gun\_ownership & 0.61 & infring & 0.87 \\
			firearm & 0.58 & constitut & 0.85  \\
			firearm\_ownership &  0.56 & amend & 0.83 \\
			religi &  0.54 & enshrin & 0.81  \\
			ideolog &  0.54 & regul\_militia & 0.80   \\
			advoc &  0.54 & inalien & 0.78 \\
			facet &  0.53 & free\_speech & 0.76  \\
			bear\_arm &  0.52 & constitution & 0.75  \\
			principl &  0.51 & ammend & 0.75  \\
			belief & 0.51 & god-given & 0.74 \\
			\hline
		\end{tabular}
		\caption{Top 10 similar words to religious-related keywords resulted from Word2Vec.}
		\label{tab:religion}
	\end{table}
	\begin{table}[t]
		\begin{tabular}{l c | l c }
			\hline
			\textbf{Key Word} & \textbf{Cosine} & \textbf{Key Word} & \textbf{Cosine} \\
			(self\_defense) &\textbf{Sim.} & (evil+firearm)&\textbf{Sim.}  \\
			\hline
			lethal\_forc & 0.63 & oppress & 0.68 \\
			fear\_life & 0.62 & disarm & 0.66  \\
			deadli\_forc &  0.60 & tyrant & 0.65 \\
			innoc &  0.58 & populac & 0.63  \\
			mortal &  0.58 & defend\_themselv & 0.62   \\
			merci &  0.57 & rebellion & 0.62 \\
			divin &  0.57 & tyranni & 0.61  \\
			evil &  0.57 & protect\_themselv & 0.60  \\
			prophet &  0.57 & polit\_societi & 0.60  \\
			justic & 0.57 & martyr & 0.60 \\
			\hline
		\end{tabular}
		\caption{Top 10 similar words to self-defense-related keywords resulted from Word2Vec.}
		\label{tab:religion_2}
	\end{table}
	\begin{table}[t]
		\small
		\centering
		\begin{tabular}{l c |  l r}
			\hline
			\textbf{Key Word} & \textbf{Cosine} & \textbf{Key Word} & \textbf{Cosine} \\
			(freedom) & \textbf{Sim.}& (patriot+gun)&\textbf{Sim.} \\
			\hline
			liberti & 0.69 & citizen &0.57 \\
			privileg & 0.68 &    politician & 0.57  \\
			bear\_arm & 0.67 &   freedom &0.57  \\
			freedom\_speech & 0.65 & gun\_owner &0.56 \\
			godgiven & 0.63 & american & 0.55\\
			tyrant & 0.63 &  gun\_ownership& 0.54\\
			inalien & 0.62 &  govern & 0.53\\
			sovereignti & 0.62 & tyranni & 0.52 \\
			oppress & 0.61 &  countri & 0.51 \\
			patriot & 0.60 & liberti & 0.50 \\
			\hline
		\end{tabular}
		\caption{Top 10 similar words to freedom and patriot keywords resulted from Word2Vec.}
		\label{tab:freedom_patriot}
	\end{table}

	\subsubsection{\bf Sentimental Feelings and Group Identity} 
	Gun ownership is often motivated by the significant value that guns hold for individuals and groups, serving as markers of identity within families and social circles.
	Guns can also be treated as collectibles with sentimental value for their owners, and are often passed down through generations.~\cite{texasfirearms22, yamane2017sociology, witkowski2014visual}.
	We next delve into these attachments by examining concepts related to gun collections and their social values.

	{\bf Gun Collection} We investigate gun collection via our model using the keywords ``gun+collect,'' resulting in similar words listed in Table~\ref{tab:collection_social}, including ``safe\_queen,'' ``hobbi,'' ``antiqu,'' ``sentiment\_valu,'' and ``heirloom.''
	The term ``safe\_queen'' is slang for firearms prized solely for their collectibility rather than utility (i.e., they are kept locked up in a safe as opposed to being fired).
	Phrases like ``sentiment\_valu`` and ``heirloom'' suggest that firearms hold emotional significance and value beyond their practical or monetary worth, often passed down through generations within families.
	The presence of terms like ``milsurp'' (Military surplus) and ``handgun'' among similar words indicates the types of goods and firearms collectors are primarily interested in acquiring.

	{\bf Social Value} Firearms can shape individual identities and social circles, to the extent that not owning one can lead to stigma~\cite{kalesan2016gun}. Exploring similar words to ``gun+friend+famili'' yields terms like ``commun,'' ``encourage,'' ``social\_circle,'' and ``motiv,'' (see Table~\ref{tab:collection_social}).
	This is the indication of social value of firearms that promotes social bonds and a sense of community among users on 4chan's /k/, highlighting the value of firearms beyond their mechanical worth for individuals and families.
	
	\subsubsection{\bf Masculinity} Firearms, often associated with power, symbolize masculinity, as observed in rituals like gifting boys firearms upon reaching puberty~\cite{yamane2017sociology}. This symbolism extends to hunting, providing avenues for men to express skills, dominance, and competition~\cite{hunting_masculinity}.
	In exploring Gun Culture 1.0, we focus on marksmanship and hunting activities to understand how users on 4chan express power through masculinity and dominance.

	{\bf Marksmanship} Our model's top 10 most similar words to ``marksmanship'' include ``proficiency,'' ``competitive\_shooter,'' and ``long\_range'' as shown in Table~\ref{tab:marksmanship}. 
	All of these terms represent various expressions of masculinity, as discussed by Littlefield~\cite{hunting_masculinity}. 
	
	{\bf Hunting} In the list of top 10 similar words to ``hunt'' we see words like ``small\_game,'' ``medium\_game,'' and ``danger\_game,'' which refer to the size of the targeted hunt. 
	For example, animals in ``danger\_game'' include lions, buffalos, large bears, etc.~\cite{hunting_games}.
	By examining similar words to ``hunt,'' we also find the term ``trophi\_hunt'' within the top 50 with a cosine similarity of 0.52.
	Therefore, we investigate our model with ``trophi\_hunt'' as well.
	This yields terms like ``bear\_wolv'' examples of wild animals that are not submissive to human authorities representing power and dominance~\cite{manenough}.
	 
	\subsubsection{\bf Moral Empowerment}
	Guns offer not only physical empowerment but also moral and emotional empowerment.
	Bearing arms is intertwined with religious beliefs; a perceived ``God-given'' right enabling gun owners to face and defend against evil forces~\cite{texasfirearms22}.
	This includes the perceived need for armed patriots to defend freedom against a tyrannical government and evil forces as gun ownership is perceived a crucial aspect of being a ``good and patriotic American~\cite{mencken2019gun}.''
	In our investigation, we explore these keywords to understand gun attachment from moral and religious perspectives.
	
	{\bf Religion} 
	Drawing from~\cite{texasfirearms22} and Gun Culture 2.0, we explore the intersection of gun culture and religion with the keyword ``gun+religion'' (Table~\ref{tab:religion}). ``Bear\_arm,'' central to Second Amendment interpretation, is linked to terms like ``infring,'' ``amend,'' ``free\_speech,'' and ``god-given,'' highlighting the religious perspective of gun rights organizations discussed by users~\cite{texasfirearms22}.
	
	In Gun Culture 2.0, self-defense is viewed as a divine right against threats, reflected in the keyword ``self\_defense'' (Table~\ref{tab:religion_2}). Associated terms like ``evil,'' ``disarm,'' and ``tyranny'' reveal discussions about defending freedom from oppressive regimes and the perceived necessity of gun ownership for this purpose among users discussions~\cite{texasfirearms22}.

	{\bf Freedom and Patriotism}
	Interpretations of the Second Amendment, focusing on self-defense~\cite{texasfirearms22, yamane2017sociology}, intertwine gun ownership with patriotism, actively promoted by the NRA~\cite{dawson2019shallNRA}. This connection forms the basis for gun attachment, explored through terms similar to ``freedom'' and ``patriot+gun'' (Table~\ref{tab:freedom_patriot}). Terms like ``bear\_arm,'' ``godgiven,'' and ``tyrant,'' align with the NRA's portrayal of guns as symbols of freedom and patriotism. The term ``godgiven'' reveals religious undertones in these concepts.

\begin{figure}[t]
		\centering
		\includegraphics[width=1.0\columnwidth]{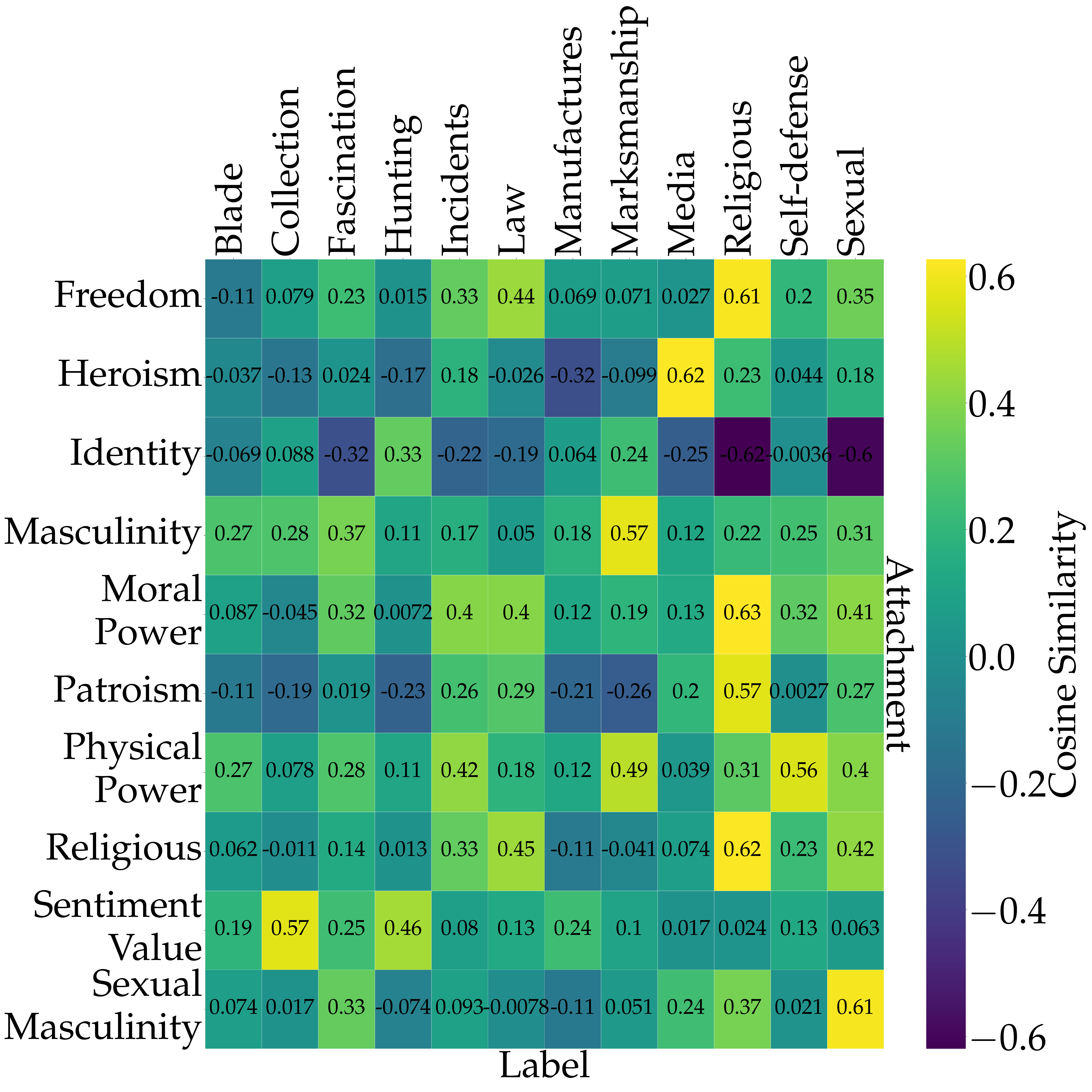}
		\caption{Similarity between different types of gun attachment and discussion labels.}
		\label{fig:gun_attachment}
	\end{figure}

	We summarize different types of gun attachment and their association with discussion labels in Figure~\ref{fig:gun_attachment}. To create this matrix, we calculate the cosine similarity score between the average of the most similar vectors to the discussion labels and the vectors corresponding to the keywords of each attachment. For instance, the ``Physical Power'' attachment is more prominent in discussions labeled ``Self-defense'' and ``Marksmanship,'' while Patriotism is more prevalent in discussions categorized under ``Law'' and ``Religious'' labels.

	{\bf Takeaways.} 
	Using the Word2Vec model to analyze keywords related to Gun Culture 1.0 and 2.0, along with motivations for gun attachment, allows for a nuanced exploration of these concepts in user discussions. Our analysis unveils the context and extent of discussions around different type of attachments to guns, revealing that users express concepts of Gun Culture 1.0 like masculinity, dominance, and sentimental value through activities like hunting, competitions, and social circles.
	When exploring the use of language with Word2Vec, we also shed light on concepts related to Gun Culture 2.0 and attachments like moral empowerment, intertwined with freedom, patriotism, and religious beliefs, suggesting subtle expressions of gun fetishism or obsession, which we explore further in subsequent sections.

	\section {Gun Fetishism}
	\begin{table}[t]
		\small
		\centering
		\begin{tabular}{l c | l c }
			\hline
			\textbf{Key Word} & \textbf{Cosine} & \textbf{Key Word} & \textbf{Cosine} \\
			(fetish) & \textbf{Sim.}& (fetishist+gun)&\textbf{Sim.}  \\
			\hline
			pervert & 0.78 & nongun & 0.55  \\
			tomboi & 0.75 & feel\_emascul & 0.54  \\
			coomer & 0.74 &obsess & 0.53  \\
			futa & 0.74 & nogun & 0.52  \\
			fetishist & 0.73 & gunfag & 0.51  \\
			bisexu & 0.72 & incel & 0.50  \\
			arous & 0.72 & fetish & 0.50  \\
			feminin & 0.72 & inanim\_object & 0.50  \\
			anim\_girl & 0.71 & normalfag & 0.50 \\
			bondag & 0.71 &revolverfag &0.49  \\
			femboi & 0.71 & enthusiast & 0.49 \\
			deviant & 0.71 & swoon & 0.48 \\
			lesbian & 0.70 & nofun & 0.48 \\
			cute\_girl & 0.70 & scary & 0.47 \\
			short\_hair & 0.69 & sexual\_deviant & 0.46  \\
			lust & 0.68 & glockfag & 0.46 \\
			\hline
		\end{tabular}
		\caption{Top 15 most similar words to fetishistic keywords resulted from Word2Vec model.}
		\label{tab:fetish}
	\end{table}
	
	\begin{table*}[t]
		\small
		\begin{tabular}{| p{0.1\linewidth}| p{0.89\linewidth} |}
			\hline
			Label & Comment   \\
			\hline
			\begin{small}Collection\end{small}&
			\begin{small}``I've been struggling with this lately actually. I'm up to 20 guns and I feel like a fucking hoarder. I can't shoot  20 guns routinely, I can barely keep up with 4 or 5. On the other hand, I'm genuinely really into all of my guns and would have a hard time selling them. I think I need to just get over my gun fetish and pare my hoard down to shit I actually shoot.''\end{small}
			\\
			\hline
			\begin{small}Maintenance\end{small} & 
			\begin{small}``Greasy looking wood on old guns is my fetish. Whether its mid 20th century hunting guns or 90s police trade ins, that glossy used look is absolute kino''\end{small}
			\\
			\hline
			\begin{small}Manufactures\end{small}& 
			\begin{small}``Yes, only because I have a fetish for failed AR-18  and other similar wierd guns. ACR, Faxon ARAK, Leader T2, etc.'',
				``spend 900 dollars on a handgun. LOOK, IM NOT POOR NOW PLS RESPECT ME!  this new wave of HK fetishism is so tiring.,'' ``small folding guns is my fetish.''\end{small}
			\\
			\hline
			\begin{small}Fascination\end{small} & 
			\begin{small}``Wound cavity fetish aside the best thing about watching people shoot that round is how flat the gun stays.,'' ``Firing a 12ga shotgun for the first time was like discovering a brand new fetish you never knew you always needed to have. Absolutely recommend for first gun.''\end{small}
			\\
			\hline
			\begin{small}Hunting\end{small}&
			\begin{small}``As long as the beast is cooked and eaten, then I am happy. I am relegated to killing minor pests. The rabbit. The wood pigeon. These creatures, although agricultural pests, and although delicious, provide little danger to the hunter. The thrill is still there, but it is fleeting, almost shameful. Like masturbating to a fetish you know is wrong, and seconds after release feeling disappointed in yourself.''\end{small}
			\\
			\hline
			\begin{small}
				Marksmanship
			\end{small} &
			\begin{small}``I don't know of a way other than practice. My impression is that I don't get better from 200 rounds every few weekends. [...] I am considering getting some sort of pellet gun to practice, but controlling recoil is huge in competitions, and that's obviously not something you get with an air/CO2 gun. [...] So if someone can recommend one. I have put more 9mm down range this summer than a lot of people here. I just don't see it as a viable way to improve unless I plan to compete or it's a fetish.''\end{small}
			\\
			\hline
			\begin{small}Sexual\end{small} & 
			\begin{small}``favorite gun AK-47, sexual fetish, Skirts and stockings''\end{small}
			\\
			\hline
		\end{tabular}
		\caption{Example of posts expressing gun fetishism.}
		\label{table:fetish_confess}
	\end{table*}
	Some attachments to guns may evolve into fetishism. In this section, we specifically analyze our dataset for signs of fetishism and explore methods to identify it in posts.

	\subsection{Confession of Fetishism}
	Kaplan defines fetishism as a widely held false belief that can manifest directly or indirectly, with some individuals openly confessing their fetish for specific objects~\cite{kaplan2006cultures}. We initially scan our dataset for terms derived from ``fetish,'' revealing 1,346 posts associating guns with fetishism or containing confessions. Examples are shown in Table~\ref{table:fetish_confess}.

	Next, we explore the most similar words to ``fetish'' from our trained Word2Vec model, revealing terms with sexual connotations, as detailed in Table~\ref{tab:fetish}. When examining words similar to ``fetishist+gun,'' we find phrases like ``gunfag,'' ``revolverfag,'' and ``glockfag,'' typical 4chan terminology describing individuals with a particular affinity for specific items. E.g., a ``glockfag'' is someone that likes, often to the point of explicitly collecting, wearing branded clothing, etc., guns produced by the manufacturer Glock.

	\subsection{Discovering Fetishism}
	In this section, we create a semantic similarity pipeline to answer our fourth research question: \emph{Is it possible to identify whether a post reflects fetishism or not?} Semantic similarity techniques model textual contexts as embedding vectors, with vector distance indicating contextual similarity.

	\subsubsection{Semantic Similarity Search}
	\begin{figure}[t]
		\centering
		\includegraphics[width=0.85\columnwidth]{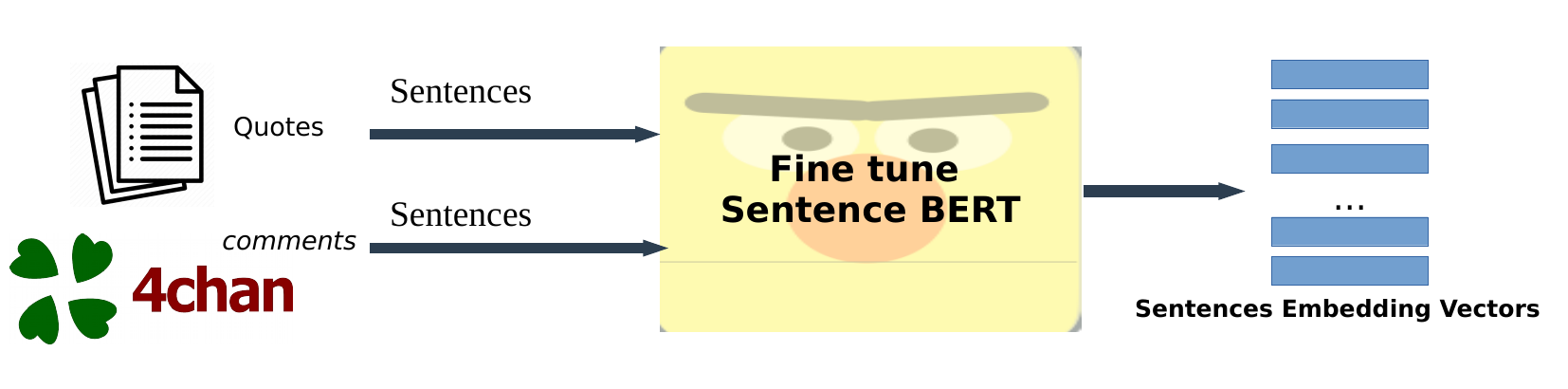}
		\caption{Sentence transformer model training process.}
		\label{fig:bert_sentence_diagram}
	\end{figure}
	\begin{table*}[t]
		\small
		\begin{tabular}{| p{0.85\linewidth}| p{0.12\linewidth} |}
			\hline
			Quote & Gun Fetishism  \\
			\hline

			``As a Christian Pastor I believe that without a deep seated belief in God and firearms that this country would not be here. I am not ashamed of that fact. I am proud of it''~\cite{springwood2014gunConcealment}& religious \\
			\hline
			
			``God created man but Sam Colt made them all equal''~\cite{texasfirearms22}&religious\\
			\hline
			``The Bible says that He who has no sword must sell his garments and buy one.''~\cite{springwood2014gunConcealment} & religious  \\
			\hline
			``America was invented by Christian progressives who all carried guns and believed that the nation was born in the context of Gods relationship to man. Jesus and religious sentiment was muscular and manly. That Christ and the nation was founded on masculine tough love that could stand strong. The Bible says that He who has no sword must sell his garments and buy one. Today I guess that would mean an AR15''~\cite{springwood2014gunConcealment} & religious  \\
			\hline

			``NRA will always defend that God given birthright from every enemy out there both foreign and domestic.''~\cite{dawson2019shallNRA}& religious \\
			\hline
			``In John Chapter 2 we see that Jesus is an assault weapons manufacturer.''~\cite{vice_moonies} & religious  \\
			\hline
			``God and guns keeps us strong. that is what this country was founded on. Well we might as well give up and run if we let them take our God and guns.''~\cite{austin2020god}& religious  \\
			\hline
			``you cant be christian if you don't own a gun.''~\cite{austin2020god} & religious \\
			\hline
			``For he is Gods servant for your good. But if you do wrong be afraid for he does not bear the sword in vain. For he is the servant of God an avenger who carries out Gods wrath on the wrongdoer.''~\cite{texasfirearms22} & religious \\
			\hline
			``I feel more moral when I carry a gun''~\cite{texasfirearms22} & religious  \\
			\hline
			``I believe I am a better person because I carry guns.''~\cite{texasfirearms22} & religious  \\
			\hline
			``The majesty of the Second Amendment that our founders so divinely captured and crafted into your birthright guarantees that no government desperate no renegade faction of armed forces no roving gangs of criminals no breakdown of law and order no massive anarchy no force of evil or crime or oppression from within or from without can ever rob you of the liberties that define your Americanism.''~\cite{dawson2019shallNRA} &  patriotism  \\
			\hline
			``second amendment is a cultural solidarity and commitment rooted in identity morality and patriotism of gun ownership.''~\cite{dawson2019shallNRA} &  patriotism \\
			\hline
			``I think the thing with the guns is more about freedom and rights.''~\cite{vice_moonies} &  freedom  \\	
			\hline	
			``stand for freedom, self reliance and the ability to control your own destiny. Gun rights are about living in a country where families are tough enough and responsible enough to stand up for themselves in a dangerous world.''~\cite{austin2020god} & freedom, empowerment  \\
			\hline
		\end{tabular}
		\caption{Examples of collected quotes linking firearms and ownership with notions of religion, patriotism, and freedom.}
		\label{tab:collected_quotes}
	\end{table*}
	Our approach starts by collecting 50 quotes demonstrating gun fetishism/obsession from sociological literature, serving as our baseline for identifying gun fetishism/obsession. The quotes are categorized under religious, patriotic, and freedom aspects of gun fetishism/obsession, with examples in Table~\ref{tab:collected_quotes}.
	
	Our system consists of two parts: Encode and Decode.
	
	\descr{Encode:}
	First, we fine-tune a pre-trained SBERT (Sentence BERT) model to identify posts closely resembling our baseline gun fetishism quotes.
	We use SBERT due to its superior performance in semantic search, compared to other BERT-based models~\cite{sbert}.
	As SBERT operates at the sentence level, we must first break down both the posts and collected quotes into separate sentences. Each sentence is identified by the presence of the characters ``.,'' ``!,'' or ``?'' at the end.
	The sentences, derived from quotes and posts serve as inputs for fine-tuning SBERT.
	We use the technique discussed in~\cite{wang2021tsdae} to fine-tune our model.
	Once SBERT is fine-tuned, we can obtain vector embeddings for each sentence from the model.
	This process is shown in Figure~\ref{fig:bert_sentence_diagram}.
	
	\descr{Decode:}
	We construct embeddings for posts and quotes by averaging their corresponding sentence embeddings.
	\subsection{Semantic Similarity Search Results}
	\begin{figure}[t]
		\centering
		
		\subfigure{\includegraphics[width=0.45\columnwidth]{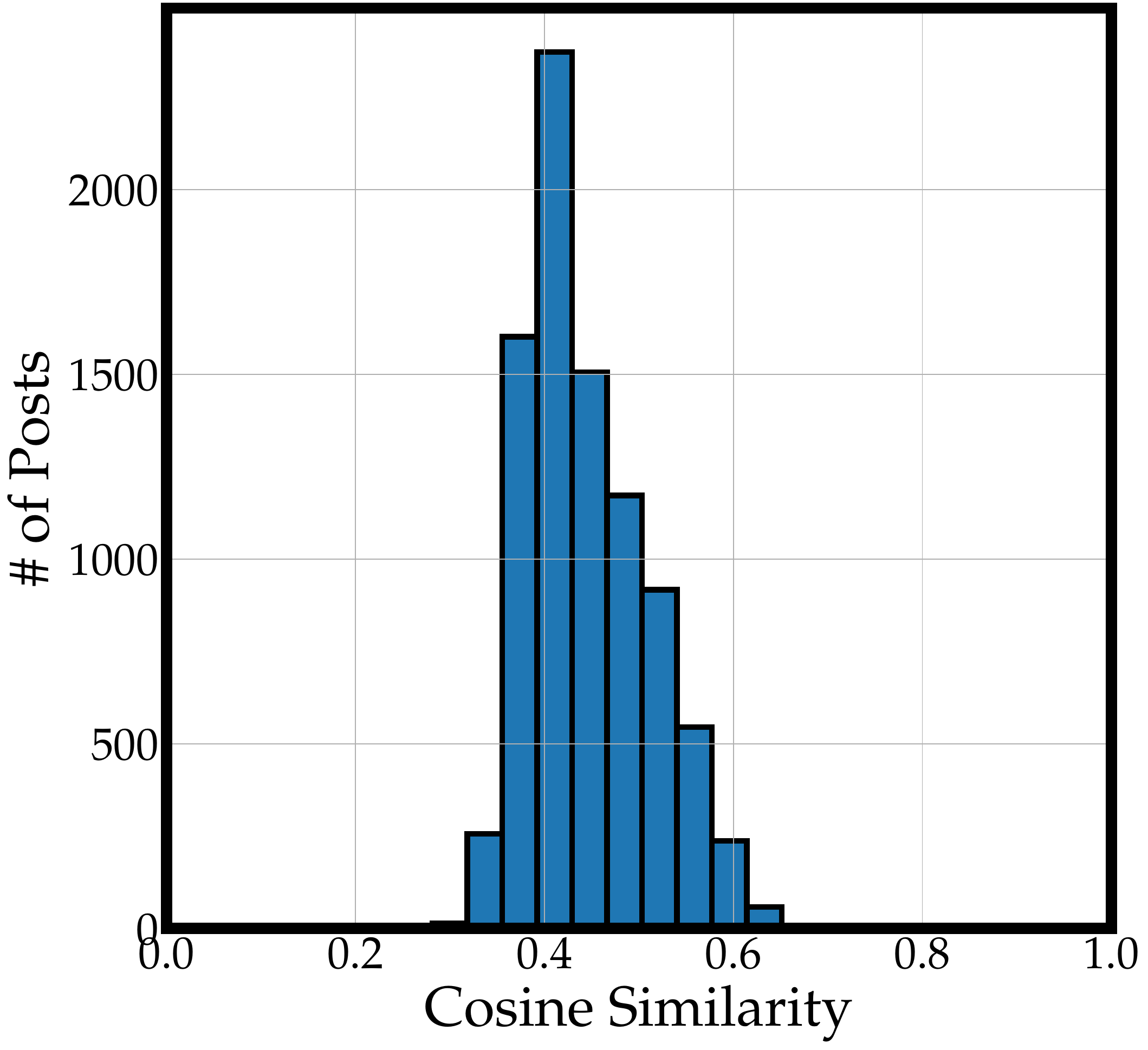}}
		\subfigure{\includegraphics[width=0.45\columnwidth]{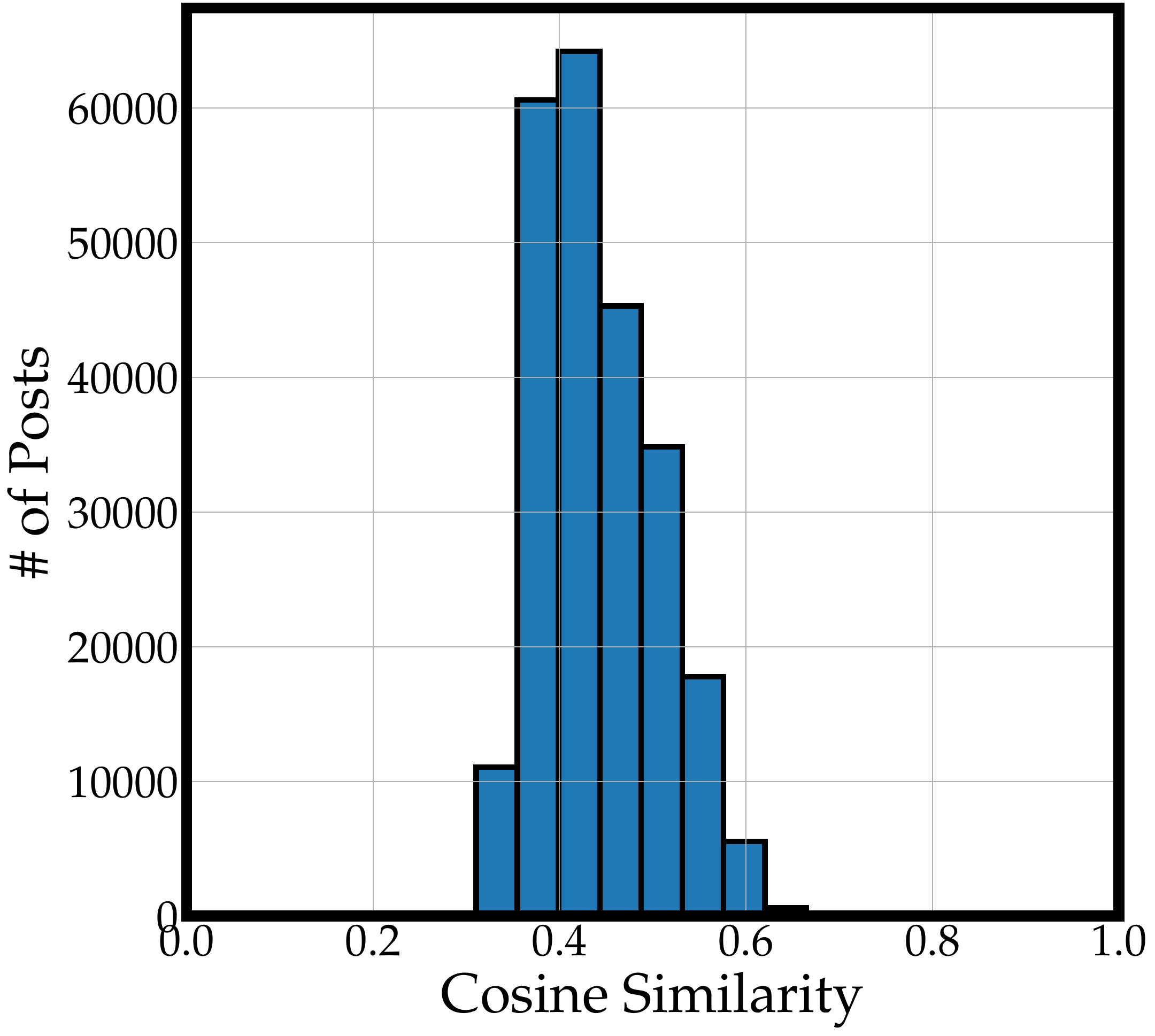}}
		\caption{Histogram of average cosine similarity between collected quotes and posts labeled (a)~Religious and (b)~Law.}
		\label{fig:law_cos_sim}
	\end{figure}
	
	\subsubsection{Religious}
	We collect quotes with religious messages supporting gun ownership from the following sources~\cite{texasfirearms22,austin2020god,dawson2019shallNRA,dipippa2014god,vice_moonies,springwood2014gunConcealment}.
	To identify semantically similar posts efficiently, we focus on those labeled as ``Religious'' in our dataset. 
	Finally, we calculate the cosine similarity score between quotes and filtered posts. The histogram of the cosine similarity score is shown in Figure~\ref{fig:law_cos_sim}.
	We then manually examine the posts with the greatest similarity scores. 
	The following post is an example resulting from the semantic similarity search which posits that Jesus would have been an advocate of gun ownership and enjoyed /k/'s culture:
	\begin{quote}
		``God is also a giant made of light sitting on a throne that surfs on a river of fire served by giant orbiting rings of eyes and wings. [...] Christ (if he ever existed, and if the description in the bible resembles him at all) was very definitely not a fan of rendering any more to Caesar than that which was Caesars, and he pretty explicitly advocated for personal ownership of weapons for everyone. Jesus was pretty /k/. [...]''
	\end{quote}
	Another user self-identifies as a ``Christ poster'' and suggests that people should sprinkle holy water on their rifles to bless them, which clearly demonstrates religious fetishism in gun ownership:
	\begin{quote}
		``Hello I am the Christ poster. I would just like to say this thread OP is based and I'm full of joy seeing how you all agree that this is Christian territory.[...] You are strong in your Armour of God, the devil has no power over you! [...] Also buy a T81 and put some holy water on your finger and then made the sign of the cross on the stock, body, handguard. I saw a Slav orthodox priest soaking rifles in Holy Water, so I thought it was an excellent idea as a layman can ask God to bless anything for us, and the sacramental of Holy Water is a beautiful method of expelling all chink evil from the wood and metal. [...]''
	\end{quote}
	Several posts underscore the idea of ``God-given'' rights to possess weapons for self-defense. For instance:
	\begin{quote}
		``You need Christ. [...] The Jews have lost their chance at salvation. This is why they are a disgusting people who hate life. They hate natural law, they hate God, they hate weapons, self defense, and the continuation of life.  God Given rights to them are nothing, [...].''
	\end{quote}
	There are also posts which elevate gun fetishism by treating firearms as objects of worship.
	For example, there are several posts with users praising a sculpture of guns, which forms a massive cube. 
	This sculpture is referred to as ``Murder/k/ube,'' and is worshipped by ``/k/ommandos''\footnote{https://amagicalplace.fandom.com/wiki/Murder/k/ube}. 
	Ironically, this art installation, which features 7,000 donated guns from around the world was created to challenge cultures of violence\weburlfootnote{http://www.gunsculpture.com/about}.
    The image of the gun sculpture originally appeared as a meme on /k/ as a joke\weburlfootnote{https://knowyourmeme.com/memes/sites/k--4}.
	However, eventually some individuals took it a bit more seriously and even made pilgrimages to the art sculpture to pray to it, giving the meme a life of its own.

	\subsubsection{Patriotism}
	We collected law-related quotes from various studies~\cite{dawson2019shallNRA,vice_moonies,springwood2014gunConcealment}, exemplified in Table~\ref{tab:collected_quotes}. These quotes often intertwine patriotism with the Second Amendment. To find similar posts, we concentrate on those labeled as ``Law'' in our dataset. The histogram of the cosine similarity score is shown in Figure~\ref{fig:law_cos_sim}.
	
	By manually examining similar law-related posts suggesting gun fetishism/obsession, we find notable examples.
	For example one poster says:
	\begin{quote}
		``Guns are your inalienable right, and the deep state traitors of the NWO want to take it away.  Biden is deep state. DeSantis is deep state. Zelensky/Ukraine is deep state.  Do not support these entities, for they only seek our destruction. True American patriots are here, armed with guns to resist tyrannical takeovers.''
	\end{quote}
	And another says:
	\begin{quote}
	``And there's always a tug of war between government and its citizens. Guns are powerful leverage against government tyranny. The more armed your civilians are the more secure they are against authoritarianism.''
	\end{quote}
	
	These posts assert that firearm ownership is an explicit display of patriotism; crucial to resisting tyrannical systems aimed at confiscating their weapons.
	As another example illustrating the close connection between gun ownership and the concept of freedom consider the following from what appears to be a Canadian user:
	\begin{quote}
		``realize there are gun owners and people who still love freedom in areas east of the Manitoba line, and I feel sorry for all of you [...].''
	\end{quote}

	\section{Conclusion}
	In this paper, we explored gun culture and fetishism on 4chan's /k/ through various topic and language modeling techniques. 
	We uncovered diverse cultural and moral motivations for gun ownership from /k/ users, some of which border on gun fetishism.
	Notably, Gun Culture 2.0 motivations, particularly self-defense, outweighed Gun Culture 1.0 in user discussions, while also exhibiting signs of religious backed fetishization of guns. 

	\descr{Implications.}
	The focus of /k/'s users on legal concerns has important societal implications.
	Although /k/ is dominated by Americans, users purporting to be from other countries with much stricter gun control laws are present and active in these legal discussions as well.
	This has clear implications for policymakers: gun culture aficionados are heavily engaged in discussion of law.
	While it is unlikely that fanatics will ever be convinced, understanding the nuances of /k/ users' discussion of law might help when it comes to communicating the goals and effects of various gun control efforts.

	Our findings also have relevance when it comes to understanding online extremism and violence at large.
	4chan and other fringe online communities are well established as vectors of radicalization that has too often resulted in real-world violence.
	Our work indicates that gun culture is fully entrenched within these fringe online communities, and we suspect it plays a large and important role in the radicalization process.
	This implies that deradicalization efforts and content moderation systems might benefit from specifically incorporating, e.g., signs of gun fetishization.

	\descr{Limitations.}
	Like any research, ours is not without limitations.
	In particular, we look at only one online community, and a fringe one at that.
	There are a wide variety of online, gun oriented communities and our findings from /k/ are unlikely to completely generalize to other communities.
	Further, our use of quantitative NLP techniques might miss out on particular nuances of discussion, e.g., we do not specifically distinguish things like the use of sarcasm.
	We believe that addressing both of these issues is a fruitful avenue for future work.

	%\bibliography{refs}

	\section{Ethics Statement}
	All results presented in this work are aggregate estimates and do not contain any information pertaining to individual participants.
	It is important to note that users posting on /k/ were fully aware of the public nature and free accessibility of the content they posted, as 4chan as a whole is a publicly accessible site.
	Furthermore, users on 4chan's boards, including /k/, post anonymously, making it highly improbable to ascertain their true identities.
	As such, this work is not considered human subject research by our IRB.
	Finally, we note that scientific output, especially on topics with high societal impact, can often be misinterpreted (sometimes deliberately) on social media~\cite{yudhoatmojoUnderstandingUseEPrints2023}.
	With that in mind, we emphasize that the objective of this work is not to stigmatize or classify classify all gun owners as gun fetishists.

	\appendix\section{Appendix}
	\begin{table}[h]
		\small
		\begin{tabular}{| p{0.20\linewidth}| p{0.75\linewidth} |}
			\hline
			Label & Top2Vec topics  \\
			\hline
			Manufactures &glock, longstroke, delton, solgw, mossberg,... \\
			\hline
			Law &legislative, overturn, gunlaws, amendment,...  \\
			\hline
			Sexual &women, horny, submissive, fetish, masculine,...  \\
			\hline
			Accessory &holosun, reticles, dots, holster, waistband,...\\
			\hline
			Media &books, memoirs, movies, hollywood,... \\
			\hline
			Fascination &penetrate, fragment, hollowpoint, velocity,...\\
			\hline
			Selfdefense &intruders, entrances, alarm, door, defense,...  \\
			\hline
			Incidents & cops, uvalde, shooting, killing, radicalize,...  \\
			\hline
			Transaction & gunbroker, payment, shortage, inflation,... \\
			\hline
			Blade &knives, blade, mora, sharpen, sword,...   \\
			\hline
			Maintenance & solvents, ballistol, oil, lubricant, grease,...   \\
			\hline
			Hunting & deer, hunt, coyote, grizzly, bear, cougar, cubs,... \\
			\hline
			Marks-manship & fundamentals, bullseye, practice, target, disciplines, competitions,...\\
			\hline
			Religious &kube, nex, alea, worship, bible, christ,...  \\
			\hline
			Collection &collect, sentimental, vintage, heirloom,...  \\
			\hline
		\end{tabular}
		\caption{Examples of Top2Vec topics within each annotated labels.}
		\label{tab:top2vec1}
	\end{table}
	\appendix\subsection{Word2Vec Modeling Details}
	Prior to starting our NLP analysis, it is essential to pre-process the collected comments and posts. This pre-processing involves multiple steps, such as removing Unicode characters (punctuation, emojis, and special symbols), removing links and URLs, eliminating excessive white spaces, normalizing text, removing stop words, and ultimately stemming all remaining words to break them down to their roots.
	We train our Word2Vec model on cleaned and stemmed posts of threads extracted from stage 2 of our topic analysis pipeline with an embedding size of 125, a window size of 7, and a minimum count of 30.
	
	\appendix\subsection{BERTopic Training Details}
	Prior to applying BERTopic, several preliminary steps are taken to enhance the quality of the results.
	We train a FastText model on threads and subsequently use it as the input embedding model for BERTopic.
	FastText is an extension of Word2Vec model that is efficient text classification and text learning tasks~\cite{fasttext}. 
	We use the FastText model because it can handle lengthy threads, provides word-level knowledge, and is not limited by token count constraints of transformers.
	We train with an embedding size of 300, a minimum word frequency of 50, a context window size of 7, and a skip-gram value of 1.
	Default values are used for the remaining model parameters.
	To minimize the impact of stop-words, we adopt BERTopic's recommended CounterVectorizer technique.
	
	\appendix\subsection{Top2Vec Topics for Within Each Discussion Labels}
	Table~\ref{tab:top2vec1} shows examples of topics within each discussion labels annotated in Topic Analysis section.

\end{document}